\documentclass[journal=acs photonics,manuscript=article]{achemso}

\usepackage[version=3]{mhchem} % Formula subscripts using \ce{}
\usepackage{color} % Formula subscripts using \ce{}

\author{Aobo Chen}
\author{Francesco Monticone}
\affiliation[Unknown University]
{School of Electrical and Computer Engineering, Cornell University, Ithaca, New York 14853, USA}
\email{francesco.monticone@cornell.edu}

\title[An \textsf{achemso} demo]
  {Dielectric nonlocal metasurfaces for fully solid-state ultra-thin optical systems}
  
\begin{document}

%%%%%%%%%%%%%%%%%%%%%%%%%%%%%%%%%%%%%%%%%%%%%%%%%%%%%%%%%%%%%%%%%%%%%
%% The "tocentry" environment can be used to create an entry for the
%% graphical table of contents. It is given here as some journals
%% require that it is printed as part of the abstract page. It will
%% be automatically moved as appropriate.
%%%%%%%%%%%%%%%%%%%%%%%%%%%%%%%%%%%%%%%%%%%%%%%%%%%%%%%%%%%%%%%%%%%%%
\begin{tocentry}
\includegraphics[ width=1\linewidth, keepaspectratio]{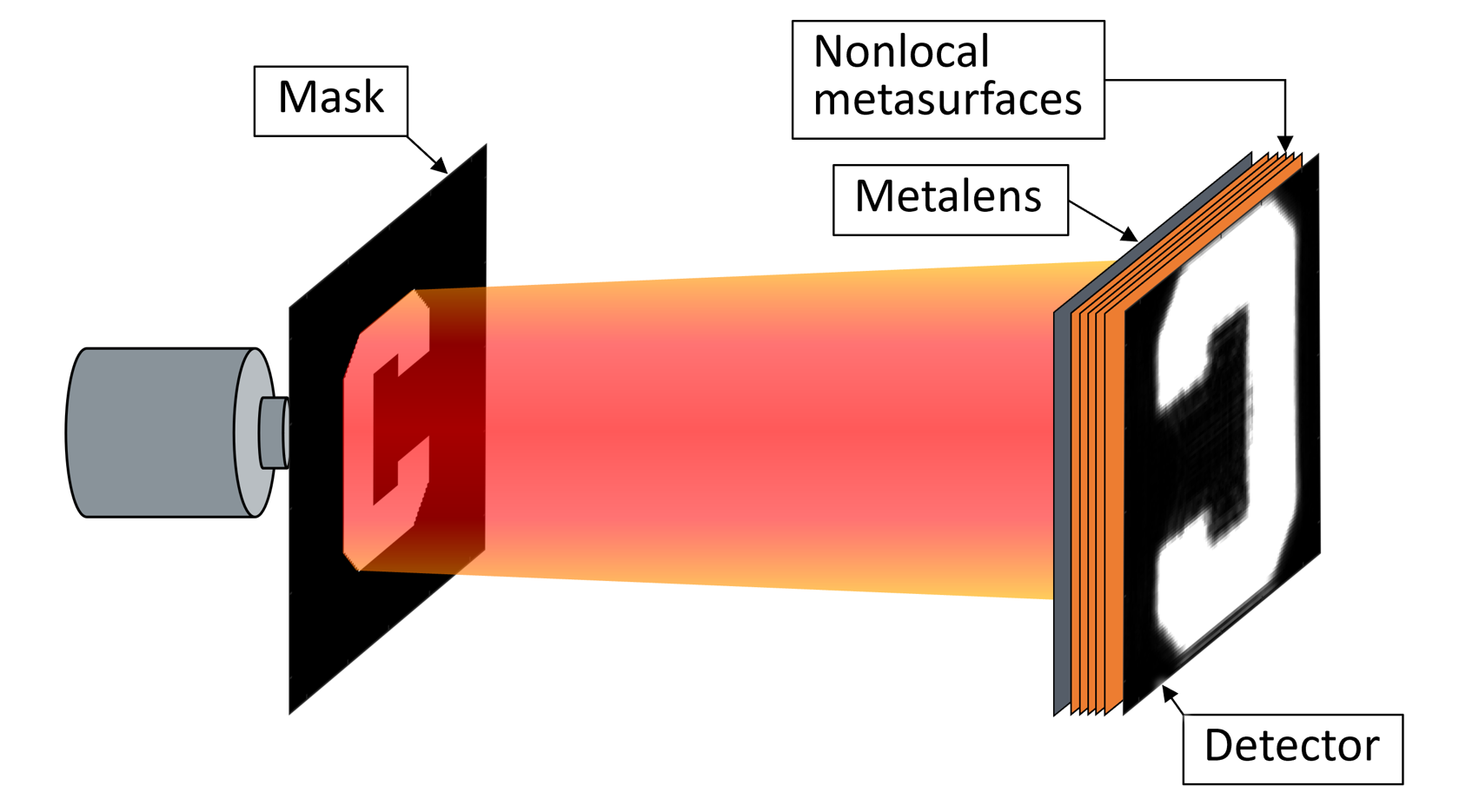}

\end{tocentry}

%%%%%%%%%%%%%%%%%%%%%%%%%%%%%%%%%%%%%%%%%%%%%%%%%%%%%%%%%%%%%%%%%%%%%
%% The abstract environment will automatically gobble the contents
%% if an abstract is not used by the target journal.
%%%%%%%%%%%%%%%%%%%%%%%%%%%%%%%%%%%%%%%%%%%%%%%%%%%%%%%%%%%%%%%%%%%%%
\begin{abstract}
  Most optical systems involve a combination of lenses separated by free-space regions where light acquires the required angle-dependent phase delay for a certain functionality. Very recently, flat-optics structures have been proposed to compress these large free-space volumes and miniaturize the overall optical system. However, these early designs can only replace free-space volumes of limited length, or operate in a very narrow angular range, or require a high-index background. These issues raise questions about the applicability of these devices in practical scenarios. Here, we first derive a fundamental trade-off between the length of compressed free space and the operating angular range, which explains some of the limitations of earlier designs, and we then propose a solution to relax this trade-off using nonlocal metasurface structures composed of suitably coupled resonant layers. This strategy, inspired by coupled-resonator-based band-pass filters, allows replacing free-space volumes of arbitrary length over wide angular ranges, and with very high transmittance. Finally, we theoretically demonstrate, for the first time, the potential of combining local and nonlocal metasurfaces to realize compact, fully solid-state, planar structures for focusing, imaging, and magnification, in which the focal length of the lens (and hence its magnifying power) does not dictate the actual distance at which focusing is achieved. Our findings are expected to extend the reach of the field of metasurfaces and open new unexplored opportunities.
\end{abstract}


%%%%%%%%%%%%%%%%%%%%%%%%%%%%%%%%%%%%%%%%%%%%%%%%%%%%%%%%%%%%%%%%%%%%%
%% Start the main part of the manuscript here.
%%%%%%%%%%%%%%%%%%%%%%%%%%%%%%%%%%%%%%%%%%%%%%%%%%%%%%%%%%%%%%%%%%%%%
\section{Introduction}
Optical systems aim to control the flow of light for different applications, such as imaging, spectroscopy, sensing, light concentration, etc. Lenses, which are the essential component of most optical systems, control light by locally molding the phase of a propagating wave through its interaction with a transversely inhomogeneous structure. Alternatives to conventional curved lenses have been investigated for centuries to miniaturize optical systems and/or achieve better optical performance (a notable example is the Fresnel lens) \cite{Born1999}. However, drastic progress in lens miniaturization has only been achieved recently with the advent of the field of flat optics and the introduction of increasingly advanced metasurfaces and metalenses \cite{Lalanne2017,Yu333,Zhao2012,Huang2012,PhysRevLett.110.203903,PhysRevLett.110.197401,Lin298,Arbabi2015,Khorasaninejad2016,Wang2018,Chen2018,Khorasaninejadeaam8100,Tseng2018,banerji2019imaging,Engelberg,Chen2020,Presutti2020}. Regardless of the specific design, the local phase profile that a metalens needs to implement to realize focusing is
\begin{equation}\label{eq:1}
    \varphi \left( r \right)=-{{k}_{0}}\left( \sqrt{{{F}^{2}}+{{r}^{2}}}-F \right)+g,
\end{equation}
where $k_0$ is the wavenumber in the surrounding medium at the operating frequency $\omega_0$, $r$ is the radial coordinate in the transverse direction, $F$ is the focal length of the metalens and $g$ is a reference phase, independent of position but possibly dependent on frequency. Throughout this work we assume a $e^{-i\omega t}$ time-harmonic convention for all field quantities.

\begin{figure}[h]
	\includegraphics[ width=0.65\linewidth, keepaspectratio]{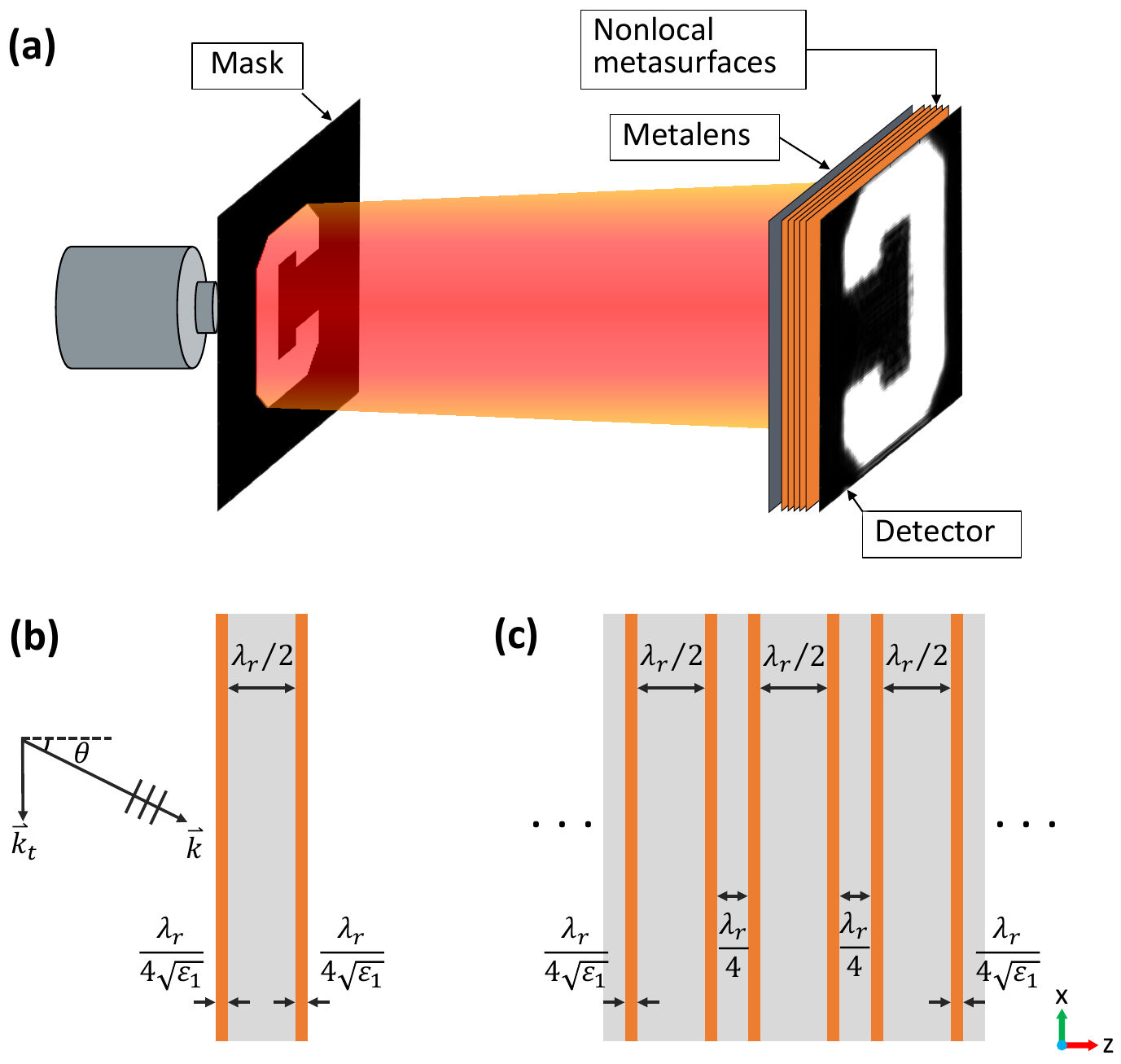}
	\centering
	\caption{\label{fig1}(a) Illustration of a ultra-thin, fully solid-state, flat-optics imaging system, in which nonlocal metasurfaces replace and compress the entire free-space volume between a flat metalens and the detector on the observation plane, hence miniaturizing the entire optical system without affecting the imaging performance. %\red{(the focal length of the metalens is unaltered, but focusing occurs at a much shorter distance from the metalens since the nonlocal metasurface implements the response of a much longer free-space region... decoupling the actual distance at which focusing is achieved from the focal length of the lens..)}
	(b) Planar dielectric structure supporting a guided-mode resonance, as an example of the simplest possible design that can implement the nonlocal response in Eq. (\ref{eq:2}). The two dielectric plates (orange) have moderately high relative permittivity $\varepsilon_1$ and are separated by a material with relative permittivity close to unity (grey). Relevant dimensions are shown in the figures, where $\lambda_r$ is the free-space wavelength at the resonant frequency, $\omega_r$, of the first even Fabry-Pérot-like resonance of the structure. (c) $n$-layer structure acting as a nonlocal metasurface, in which individual resonators are separated by quarter-wavelength spacers to obtain approximately uniform transmission amplitude and linear phase with respect to frequency, as discussed in the text. }
\end{figure}

While the field of metasurfaces holds promise to revolutionize optics by replacing bulky conventional lenses with compact planarized devices, simply making lenses flat and thin is not sufficient to miniaturize the vast majority of real-world optical systems. In fact, a significant percentage of the volume of an optical system (for example, a microscope, a telescope, or a Fourier optical processor) is composed of free space. Empty space between lenses is needed to allow light to propagate, acquire an angle-dependent phase delay, and achieve focusing at a desired point. However, only very recently attention has started to be given towards the problem of systematically miniaturizing this empty space between solid-state optical components, e.g., lenses and detectors \cite{reshef2020towards,guo2020squeeze}.

To better understand how to approach this problem, let us consider the transfer function for plane waves propagating through a free-space volume of length $L$. While the transmission amplitude is unity for propagating waves in free space, the transmission phase is a function of the transverse wavenumber (momentum) $k_t$ as
\begin{equation}\label{eq:2}
\varphi \left( {{k}_{t}} \right)=L\sqrt{k_{0}^{2}-k_{t}^{2}} \approx k_0L-\frac{L}{2k_0}k_t^2,
\end{equation}
where the expression on the right is a Taylor expansion of the transmission phase for small $k_t$. Thus, in order to replace a free-space segment of length $L$ with a planar device of thickness $T<<L$, the designed structure should have the same optical response. Then, a ``squeezing ratio'' can be defined as $R=L/T$. Most importantly, it is clear from Eq. (\ref{eq:2}) that the required transmission phase response of such a ``space-squeezing device'' depends directly on the transverse momentum $k_t$, and not on the spatial location $r$, indicating that the response of the desired device should be ``nonlocal'' \cite{Landau1984}. 

%For a local optical device, e.g., a metalens, the transmission phase is a function of spatial location $r$, as shown in Eq.(\ref{eq:1}). Instead

To clarify the terminology further, conventional metasurfaces and metalenses are ``local'' in the sense that they control the wave transmission/reflection as a function of position (locally and pointwise) as in Eq. (\ref{eq:1}), based on an engineered \emph{transversely inhomogeneous} structure, to achieve for example beam focusing or deflection. In this regard, while huge research efforts have been devoted to metasurfaces in recent years, their operation is not fundamentally different from other transversely inhomogeneous optical devices, such as conventional lenses and gratings. %The main advantage of flat optics is that it allows realizing the same functionalities of bulky optical systems over a flat and thin platform. 
In drastic contrast with such local devices, a nonlocal metasurface is a planar device designed to control the transmission as a function of the transverse wavenumber $k_t$, based on a transversely homogeneous but \emph{longitudinally inhomogeneous} (e.g., layered) structure. In other words, since the $k_t$-dependent transmission function corresponds to the angular transfer function of the device, nonlocal flat optics aims to provide a novel approach to systematically control the angular transmission response. While three-dimensional nonlocal metamaterials, with wavevector-dependent constitutive parameters, have been studied extensively in the past, the field of nonlocal flat optics -- which was pioneered in the context of wave-based analog optical computing \cite{Silva160,PhysRevX.6.041018,Guo:18,{PhysRevLett.121.173004}} -- is still in its infancy. 

%\red{The ability to control the angular transfer function at will is crucial for a number of applications for which conventional metasurface design methodologies are not sufficient. We also would like to stress that, while the angular transfer function could be engineered with conventional Fourier optics [10], typical systems of this type (e.g., a 4F correlator) require multiple lenses, transmission masks, and free-space propagation, for an overall thickness of hundreds of wavelengths. Nonlocal meta-optics allows miniaturizing such a system to a single flat device of thickness on the order of one wavelength, i.e., a miniaturization of several orders of magnitude, as originally shown in \cite{Silva160}. [maybe not needed, or as a note]}
%
The idea of applying nonlocal flat structures to the problem of compressing empty space in optical systems was originally proposed, very recently, in Refs. \cite{reshef2020towards} and \cite{guo2020squeeze}. Specifically, Ref. \cite{reshef2020towards} demonstrated a space-squeezing effect using low-index isotropic and anisotropic slabs in a high-index background, as well as using optimized multilayered structures, whereas Ref. \cite{guo2020squeeze} considered a modified photonic-crystal slab achieving the same effect. However, these early designs suffer from fundamental and practical problems, which hinder their practical potential and applicability. The structures proposed in \cite{reshef2020towards} operate over a moderately large range of transverse wavenumbers (angular range), but can only replace a limited length of free space (in addition, the only reported experimental demonstration was based on a high-index background, making it impractical for most applications). Conversely, the photonic-crystal device in \cite{guo2020squeeze} can replace a much longer free-space volume, but the angular range of the device is very limited.

Within this context, the goal of this work is three-fold: (i) We first discuss why some of these limitations are fundamental for a broad class of nonlocal metasurfaces based on a single guided-mode resonance. Specifically, we derive a quantitative trade-off between the length of replaced free space $L$ and the operating range of transverse momentum $k_t$, which explains the limited performance achieved in previous works. (ii) We then propose a solution to relax this trade-off, based on a stack of suitably coupled dielectric resonators, realizing a nonlocal metasurface device that can replace free-space regions of arbitrary length over a wide angular range. (iii) Finally, we theoretically and computationally demonstrate, for the first time, the potential of combining local and nonlocal metasurfaces to realize fully solid-state structures for focusing and imaging, as illustrated in Fig. \ref{fig1}(a), in which the actual distance where focusing is achieved is fundamentally decoupled from the focal length of the lens (and hence its magnifying power).

\section{Fundamental Trade-offs and Nonlocal Metasurface \\Design}

\subsection{Nonlocal structures based on a single guided-mode resonance -- Operational principle and trade-offs}

In order to better understand the physics and design trade-offs of nonlocal space-squeezing devices, we first consider one of the simplest possible planar structures that can implement the nonlocal phase function in Eq. (\ref{eq:2}), by relying on a guided-mode resonance, as in Ref. \cite{guo2020squeeze}. Fig. \ref{fig1}(b) shows the structure under consideration, which is essentially a simple dielectric resonator comprised of two quarter-wavelength-thick dielectric plates (acting as mirrors with relatively large, real, reflection coefficient), spaced by half a wavelength at the resonant frequency, $\omega_r$, of the first even Fabry-Pérot-like resonance at normal incidence. The spacer layer is assumed to be made of a material with relative permittivity equal to unity (but any transparent material that provides enough contrast with the dielectric plates would also work, as further discussed below). Under oblique plane-wave incidence with transverse wavenumber $k_t=k_0 \sin \theta$, at a frequency near $\omega_r$ the transmission coefficient can be described locally by a single-resonance model (see supporting information, Eq. (S5))
\begin{equation}\label{eq:3}
t\left( \omega ,{{k}_{t}} \right)= \frac{\gamma_0 }{\gamma_0 -i\left( \omega -\omega \left( k_t \right) \right)},
\end{equation}
where $\gamma_0$ is the resonance linewidth, which is assumed to be invariant with $k_t$, and $\omega \left( {{k}_{t}} \right)$ is the dispersion relation of the resonant mode, namely, the dispersion relation of the guided leaky mode responsible for the Fabry-Pérot-like resonance.
The transmission phase at frequencies near the eigenfrequency $\omega \left( {{k}_{t}} \right)$ of the guided mode is
\begin{equation}
\begin{aligned}
\arg \left( t \right)=\arctan  \frac{\omega-\omega\left( k_t  \right)}{\gamma_0}  \approx \frac{\omega-\omega\left( k_t  \right)}{\gamma_0}. 
\end{aligned}
\label{eq:linear}
\end{equation}
from which we see that the near-resonance transmission phase is approximately a linear function with respect to $\omega$ or $\omega \left( k_t \right)$. Considering the exact expression in Eq. (\ref{eq:linear}), we also note that the phase varies around the resonance (varying either $\omega$ or $k_t$) in a limited range from $-\pi/2$ to $\pi/2$, as expected. This already indicates the inherent limitation of using a single guided-mode resonance to implement the phase function in Eq. (\ref{eq:2}), which generally requires an available phase range much wider than $\pi$ to replace a free-space volume of large length over a wide range of $k_t$.

% If a dielectric slab is off-resonance, or covered by anti-reflection coatings, it has the effect of pushing the focus farther, as can be understood from a simple application of Snell's Law.

To obtain closed-form expressions for the transmission phase, we assume that the guided mode supported by the structure in Fig. \ref{fig1}(b) is well confined and with little radiation leakage, which corresponds to having a scattering resonance with sufficiently high quality factor $Q$ (which in turn requires that the permittivity of the dielectric plates is sufficiently high). In this way, the imaginary part of the eigenfrequency can be neglected, and it can be shown that the dispersion relation of the considered guided mode is approximately a quadratic function for relatively small values of $k_t$ (small incidence angles) (see supporting information, Eq. (S14)): 
\begin{equation}
\omega \left( {{k_t}} \right) \approx {\omega _r} + \alpha k_t^2.
\label{eq:quadratic}
\end{equation}
An approximate expression for the coefficient $\alpha$ of the second-order term for the structure under consideration is given in the supporting information document, Eq. (S15). Combining Eqs. (\ref{eq:linear}) and (\ref{eq:quadratic}), we get
\begin{equation}
    \arg\left(t \right)\approx\frac{\omega-\omega_r}{\gamma_0}-\frac{\alpha}{ \gamma_0 }k_t^2.
    \label{eq:phase metasurface}
\end{equation}
Thus, the transmission phase of the considered structure, acting as a nonlocal $k_t$-dependent metasurface, has the same form as that of free space, given by Eq. (\ref{eq:2}). We also note that, similar to Eq. (\ref{eq:2}), Eq. (\ref{eq:phase metasurface}) only has even-order terms, which is expected since both free space and the considered structure are transversely mirror-symmetric, hence the transmission response is the same for plane waves with opposite transverse wavenumbers, $k_t$ and $-k_t$.

As noted in \cite{reshef2020towards} and \cite{guo2020squeeze}, a global $k_t$-independent transmission-phase difference between the space-squeezing device and free space is irrelevant. Thus, we neglect the zero-order terms and, comparing the coefficients of the quadratic terms in Eqs. (\ref{eq:2}) and (\ref{eq:phase metasurface}), we find that a near-resonance structure of this type can replace a portion of free space of length
%Then, if we ignore the difference of the constant terms, which represent the global phase different, as is done in \cite{reshef2020towards, guo2020squeeze}, and compare the quadratic terms in Eqs.(\ref{eq:2}) and (\ref{eq:phase metasurface}), we get the replaced length of free space
\begin{equation}
    L= \frac{2 \alpha k_0}{\gamma_0} .
    \label{eq:replaced length}
\end{equation}
A space-squeezing effect is then obtained if the length $L$ is greater than the actual thickness $T$ of the nonlocal structure. Most importantly, we can determine a fundamental limit to this effect by noting again that the transmission phase, $\arg(t)$, in Eq. (\ref{eq:linear}) is limited in the range $[-\pi/2,\pi/2]$ due to its single-resonance nature. If we then set the constant term in Eq. (\ref{eq:phase metasurface}) to $\left( \omega - \omega_r \right) / \gamma_0=\pi/2$, which corresponds to operating slightly off-resonance, the quadratic term is bounded in the range $-\pi<-\alpha k_t^2/\gamma_0<0$, provided that all higher-order terms are zero. Hence, if $k_t^2$ is outside the range $[0,\pi \gamma_0/\alpha]$, there must be higher-order terms in Eq. (\ref{eq:phase metasurface}) to guarantee that the left-hand-side remains limited in the range $[-\pi/2,\pi/2]$; therefore, in this case, $\arg(t)$ can no longer be considered quadratic with respect to $k_t$. Thus, the maximum allowed value, $k_{t,\max}$, for which the phase function can be assumed quadratic within the range $k_t \in [0,k_{t,\max}]$ is given by $k_{t,\max}^2=\pi \gamma_0/\alpha$. Finally, using Eq. (\ref{eq:replaced length}), we obtain a quantitative trade-off between the replaced length $L/\lambda_0$ and the maximum operating range of transverse momentum $k_{t,\max}/k_0 = \sin \theta_{\max}$ (maximum angular range):
\begin{equation}
    \frac{L}{\lambda_0} \cdot \left( \frac{k_{t,\max}}{k_0} \right) ^2 \leq 1.
    \label{eq:trade-off}
\end{equation}
which indicates that a larger $L$ can only be obtained at the expense of a smaller angular range. The trade-off is written as an inequality since the assumptions leading to it may not be satisfied by a generic guided-mode resonance. %in particular, the dispersion of a sub-optimal guided mode may stop being quadratic much sooner before reaching $k_{t,max}$.} \blue{(More often, the difficulty to reach the limit is because of the nonlinear phase profile in frequency response)}. 
In addition, it can be verified from the above derivation that different choices for the constant term in Eq. (\ref{eq:phase metasurface}) (different from $\left( \omega - \omega_r \right) / \gamma_0=\pi/2$), which corresponds to operating at slightly different frequencies, always lead to a narrower angular range. %Although the Fabry-Pérot-like resonance considered here is different from the guided resonance in the photonic crystal device in \cite{guo2020squeeze} or the resonances in the spaceplate metamaterial in \cite{reshef2020towards}, the trade-off still holds qualitatively. The design in \cite{guo2020squeeze} has very large length of replaced free space, so the range of transverse momentum $k_t/k_0$ is very limited, and vice versa for that in \cite{reshef2020towards}.

Although the structure considered here is different from the photonic-crystal slab in \cite{guo2020squeeze}, the trade-off still holds, because both structures rely on essentially the same mechanism, namely, a single guided-mode resonance with an approximately quadratic dispersion relation. This trade-off explains why, in the designs proposed in Ref. \cite{guo2020squeeze}, the squeezing ratio $R$ is very high, whereas the angular range is very limited. We also note that the results in Ref. \cite{guo2020squeeze} are far from the limit defined by Eq. (\ref{eq:trade-off}) because, while this design similarly uses a single-mode resonance, the resonance corresponds to a minimum of transmission, instead of a maximum, and therefore the structure needs to be operated at a frequency far from this minimum, where the available transmission-phase variation (as a function of wavenumber) is much more limited than around resonance. This leads to a poor utilization (only a few percents) of the total available phase range provided by the resonance. Moreover, while the nonlocal metamaterial in Ref. \cite{reshef2020towards} is based on optimization and not enough information is available about its guided-mode distribution, the published results suggest it also follows a qualitatively similar trade-off. In light of the theoretical results of this section, we argue that more sophisticated nonlocal space-squeezing structures, perhaps based on optimization or inverse design, would not offer any major advantage compared to the very simple structure in Fig. \ref{fig1}(b) if they are still based on a single resonance, as they would still be constrained by the same performance trade-off.

\begin{figure}[h]
	\includegraphics[ width=0.55\linewidth, keepaspectratio]{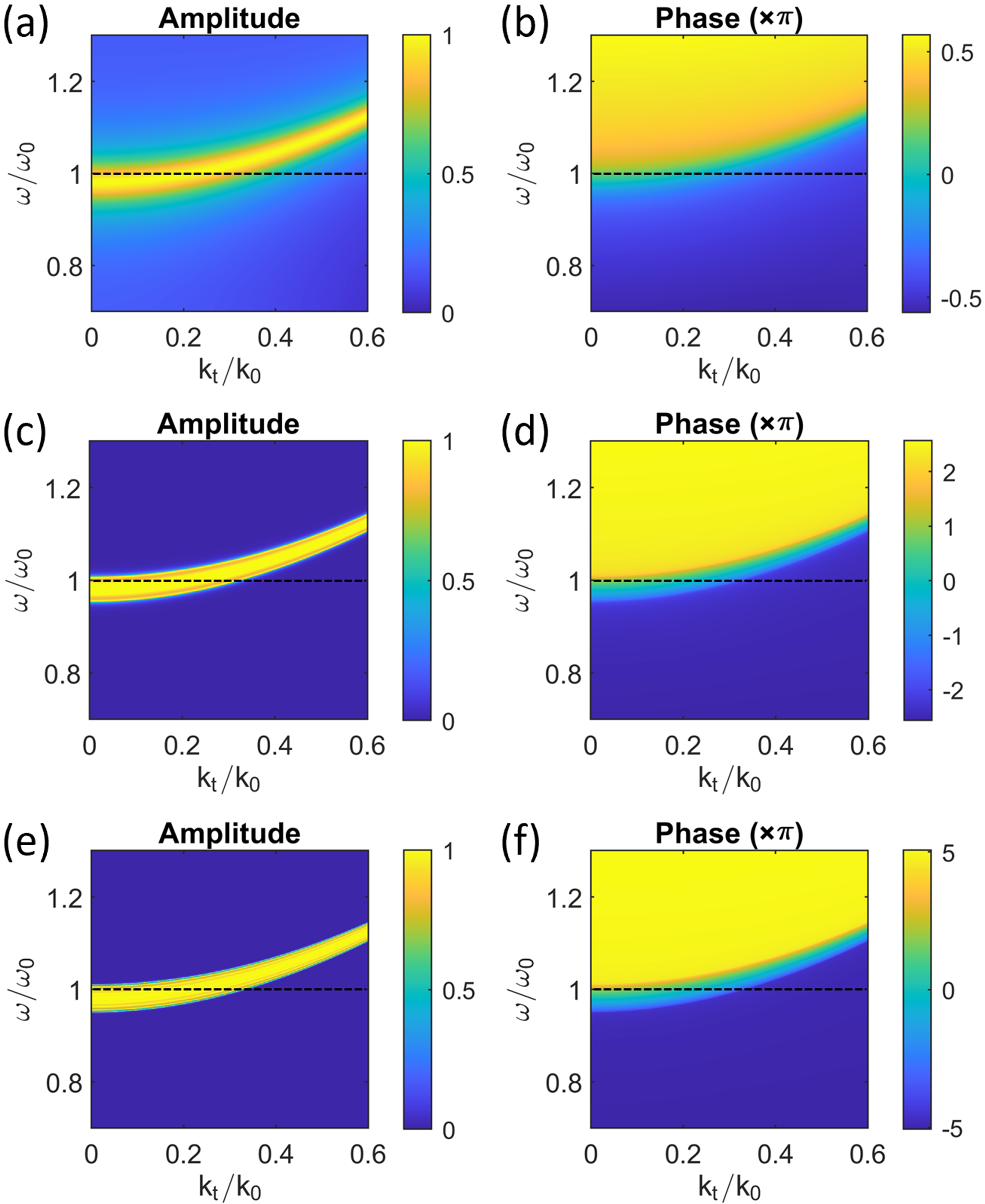}
	\caption{\label{fig2} Density plots of the transmission amplitude (left column) and phase (right column) for the proposed $n$-layer structure acting as a nonlocal metasurface (details in the text), for TE polarization, as a function of frequency $\omega$ and transverse momentum $k_t$. (a) and (b) $n=1$; (c) and (d) $n=5$; (e) and (f) $n=10$. Since the guided modes responsible for this resonance are well-confined and with low radiation loss (the eigenfrequency has small imaginary part), the bright band in the transmission amplitude plots is a good approximation of the modal dispersion relation. The horizontal dashed lines indicate the operational frequency $\omega_0$, which is chosen to be slightly off-resonance at normal incidence, i.e., $\omega_0=1.02\omega_r$, in order to utilize the widest possible angular range over which the transmission phase is a quadratic function of $k_t$.}
\end{figure}

\begin{figure}[h]
	\includegraphics[ width=0.51\linewidth, keepaspectratio]{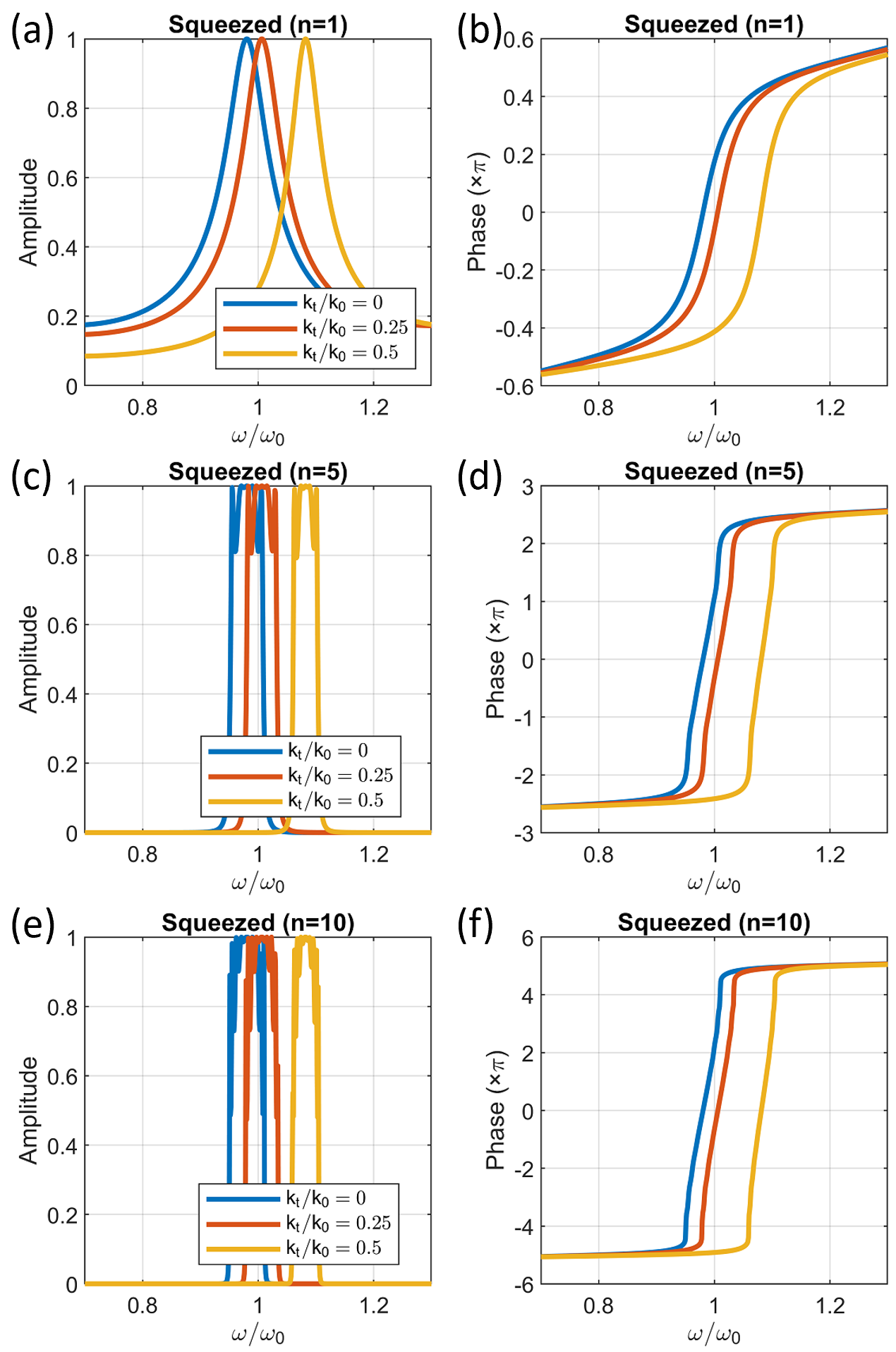}\caption{\label{fig3} Transmission amplitude (left column) and phase (right column) of the proposed $n$-layer structure acting as a nonlocal metasurface (details in the text), for TE polarization, plotted as a function of frequency $\omega$ for three different values of transverse wavenumber $k_t/k_0=0,$ 0.25 and 0.5.}
\end{figure}

\subsection{Nonlocal structures inspired by coupled-resonator-based band-pass filters}

Despite being an approximate and loose bound (due to the assumptions discussed above), the trade-off in Eq. (\ref{eq:trade-off}) is already quite stringent: as an example, a space-compression effect with $L = 5 \lambda_0$ could be achieved only over a $\pm 25$ degrees angular range, even in the best possible scenario. In order to attain significantly better performance, which would make nonlocal flat-optics systems of this type more appealing in practical scenarios, we seek to relax this trade-off between the length of replaced free space and the operating range of transverse momentum. To this aim, we need an arbitrarily large transmission phase range (not limited to $\pi$) that is linear with $\omega$ or $\omega \left( k_t \right)$ and, therefore, quadratic with respect to $k_t$ over a much broader range $[0,k_{t,\max}]$ (indeed, $k_{t,\max}$ depends on the available phase range as discussed above). Hence, the single-resonance transmission response in Eqs. (\ref{eq:3}), (\ref{eq:linear}) is clearly inadequate.

A seemingly simple solution to break the trade-off is to stack $n$ resonant structures of the type in Fig. \ref{fig1}(b), exploiting the larger total phase range of a $n$-resonator system. Indeed, Ref. \cite{guo2020squeeze} suggests considering nonlocal space-squeezing metasurfaces in cascade. However, it is crucial to note that a generic coupling between resonators would typically hinder the performance of the nonlocal structure because, while the total transmission-phase range of $n$ coupled resonators is $n\pi$, the phase function would generally not remain linear with frequency as a result of coupling-induced resonant-frequency splitting \cite{haus1984waves} (for example, for the case of two resonators of the type in Fig. \ref{fig1}(b) cascaded back-to-back with no separation, it is easy to verify that the resonant frequency of the individual resonators does no longer correspond to a transmission maximum of the two-resonator system, and the phase is no longer linear around this frequency). Hence, arranging $n$ space-squeezing devices in cascade may be used to increase the length of replaced free space, but usually at the expense of further decreasing the operating angular range, since a strongly nonlinear transmission-phase response with respect to frequency would result in a strongly non-quadratic phase response with respect to $k_t$, except perhaps in a very small range. In other words, Eq. (\ref{eq:phase metasurface}) would be valid only over a very small range of $k_t$ since the linear approximation in Eq. (\ref{eq:linear}) would be valid only over a very small range of frequencies. 

To overcome this problem, we took inspiration from conventional microwave band-pass filters, which are typically designed by suitably coupling multiple resonators in cascade \cite{Mumford1948,podgorski1980quarter,pozar2009microwave}. In particular, if several resonators (the structures in Fig. \ref{fig1}(b)) are separated by quarter-wavelength spacers, as illustrated in Fig. \ref{fig1}(c), the resulting coupling produces new resonance peaks that are suitably distributed over a certain frequency window, providing a relatively uniform in-band transmission amplitude. %(albeit with some ripples, as in so-called Chebyshev filters \cite{pozar2009microwave,podgorski1980quarter}). 
To elaborate on this point, the rationale for using quarter-wavelength spacers comes directly from linear circuit theory and transmission-line/waveguide theory. At low frequencies, where lumped-element circuit theory is a valid approximation, a band-pass maximally-flat filter can be realized by using resonant branches (lumped LC resonators) connected alternately in series and in parallel \cite{pozar2009microwave}. A configuration more suitable for high-frequency systems (e.g., microwave waveguides) is obtained by stacking parallel resonant cavities spaced a quarter wavelength apart. This design strategy makes use of the impedance-inverting property of a quarter-wave segment, which converts series-connected elements into parallel-connected elements, thereby avoiding the necessity of using both series and parallel branches \cite{Mumford1948,podgorski1980quarter,pozar2009microwave} (series elements are more difficult to implement at high frequencies). This design can then be directly translated for optical frequency operation, in the form of resonant layers separated by quarter-wavelength spacers, as in Fig. \ref{fig1}(c), to obtain the desired uniform transmission window. An approximately uniform transmission amplitude then translates into an approximately linear transmission phase within the pass-band (especially toward the center), consistent with Kramers-Kronig-like relations between amplitude and phase of the transmission coefficient of causal passive linear systems \cite{Gralak:15}.
With this strategy, therefore, one can obtain a transmission phase that covers a range of $n\pi$ and varies linearly around the resonant frequency $\omega_r$ of the individual resonators.	
%,... 1, the total phase transition of $n$ resonators $n \pi$ can be used to break the trade-off limit; 2, the shape of phase transition remains linear around the resonant frequency $\omega_r$; 3, the resonant linewidth $\gamma_0$ remains the same. 
From a simple analysis (see supporting information), it can be shown that the transmission phase of the multilayered structure around resonance ($\omega \approx \omega(k_t)$) can be approximated as
\begin{equation}
\arg \left( t \right) \approx n \cdot \frac{\omega-\omega \left( k_t \right)}{\gamma_0}. 
\end{equation}
In addition, while the coupling between resonators generates different resonance peaks covering a certain pass-band, as discussed above, the original transmission peak is still present since, exactly at resonance, each resonator looks transparent to the other resonators, and this resonance peak evolves with $k_t$ as in the single-resonator case, following the same dispersion relation (this can be seen in the examples in Figs. \ref{fig2} and \ref{fig3} discussed in the following). Thus, the transmission phase of the $n$-layer structure can be approximated as,
\begin{equation}
    \arg\left(t \right) \approx n \cdot \frac{\omega-\omega_r}{\gamma_0}-n \cdot \frac{\alpha}{\gamma_0 }k_t^2,
    \label{eq:10 n}
\end{equation}
and, by comparing again the quadratic terms in Eqs. (\ref{eq:2}) and (\ref{eq:10 n}), we find that the length of replaced free space by this $n$-layer nonlocal structure is
\begin{equation}
    L=n \cdot \frac{2 \alpha k_0}{\gamma_0} .
    \label{eq:replaced length n}
\end{equation}
Finally, following the same reasoning as before, the trade-off limit in Eq. (\ref{eq:trade-off}) becomes
\begin{equation}
    \frac{L}{\lambda_0} \cdot \left( \frac{k_{t,\max}}{k_0} \right)^2 \leq n,
    \label{eq:trade-off-n}
\end{equation}
which shows that, as we increase the number $n$ of suitably coupled elements, the product of replaced length and angular range can become arbitrarily large.

To demonstrate the potential and simplicity of this strategy to implement the optical response of free space over a much smaller length, and a wide angular range, we designed a multilayered structure acting as a nonlocal metasurface using simple dielectric plates with relative permittivity $\varepsilon_1=15$, separated by layers with relative permittivity equal to unity, as in Fig. \ref{fig1}(c). We stress that this high value for the permittivity $\varepsilon_1$ is not a requirement for the operation of the space-compression device and is considered here just as an example; materials with lower refractive index would also work, with some advantages and disadvantages. Indeed, a moderately high contrast between the dielectric layers contributes to a relatively high $Q$ factor and a small linewidth $\gamma_0$ for the Fabry-Pérot-like resonance, which determines an increase in the replaced length according to Eq. (\ref{eq:replaced length n}), but a decrease in the operating angular range due to the trade-off in Eq. (\ref{eq:trade-off}). 

The calculated transmission coefficient around the first even Fabry-Pérot-like resonances of this $n$-layer nonlocal structure is plotted, for different numbers of layers/resonators, as a function of both transverse momentum $k_t$ and frequency $\omega$ in Fig. \ref{fig2}, and as a function of $\omega$ for different values of $k_t$ in Fig. \ref{fig3}. We first consider transverse-electric (TE) polarization as an example, while transverse-magnetic (TM) can be similarly analyzed and will be discussed in the following. From these plots, we see that the total phase range of the $n$-layer structure is $n\pi$, as expected. Importantly, near resonance the transmission phase as a function of frequency can be well approximated by a linear function, whose slope linearly increases with $n$, as seen in Fig. \ref{fig3}. The transmission amplitude becomes sharper and more ``box-like'' as $n$ increases, characterized by a frequency window %\red{of width approximately equal to $\gamma_0$}, 
within which the amplitude is large and increasingly uniform, corresponding to the frequency range where the phase is approximately linear. This behavior for the transmission amplitude and phase of the proposed $n$-layer nonlocal structure near resonance is indeed analogous to that of standard band-pass filters \cite{Mumford1948,pozar2009microwave,podgorski1980quarter}. We also observe that the dispersion relation of the guided modes of the $n$-layer structure (corresponding to the bright band in Figs. \ref{fig2}(a,c,d)) has an approximate quadratic shape with respect to $k_t$, consistent with Eq. (\ref{eq:quadratic}). Note that this dispersion relation remains essentially identical as the number of layers $n$ increases since, as mentioned above, exactly at resonance each resonator looks transparent to the other resonators (its input impedance is identical to the wave impedance of the host medium) and, therefore, the original transmission peak evolves with $k_t$ as in the single-resonator case. %First, at resonance, due to the quarter-wavelength spacers, fields in nearby metasurfaces are orthogonal to each other, so resonances in different metasurfaces do not interfere with each other. Second, at resonance, the wave impedance seeing through a resonator is identical to that of free space $\eta_0$). Therefore, the physics in a metasurface at resonance is the same, regardless $n$.
%\red{Therefore, at a fixed frequency, as the transverse momentum $k_t$ increases linearly, the transmission phase of metasurfaces increase quadratically, with a factor of $n$. When $k_t$ is large, the transmission phase saturates, limited by the total phase transition of $n$ resonators, although the dispersion relations remain in the quadratic shape. Therefore, to achieve replacement of free space of a large length within a large operating range of transverse momentum, we need to stack many layers of nonlocal metasurfaces with a space of quarter wavelength between each other. [maybe move later]}

\begin{figure}[h]
	\includegraphics[ width=0.51\linewidth, keepaspectratio]{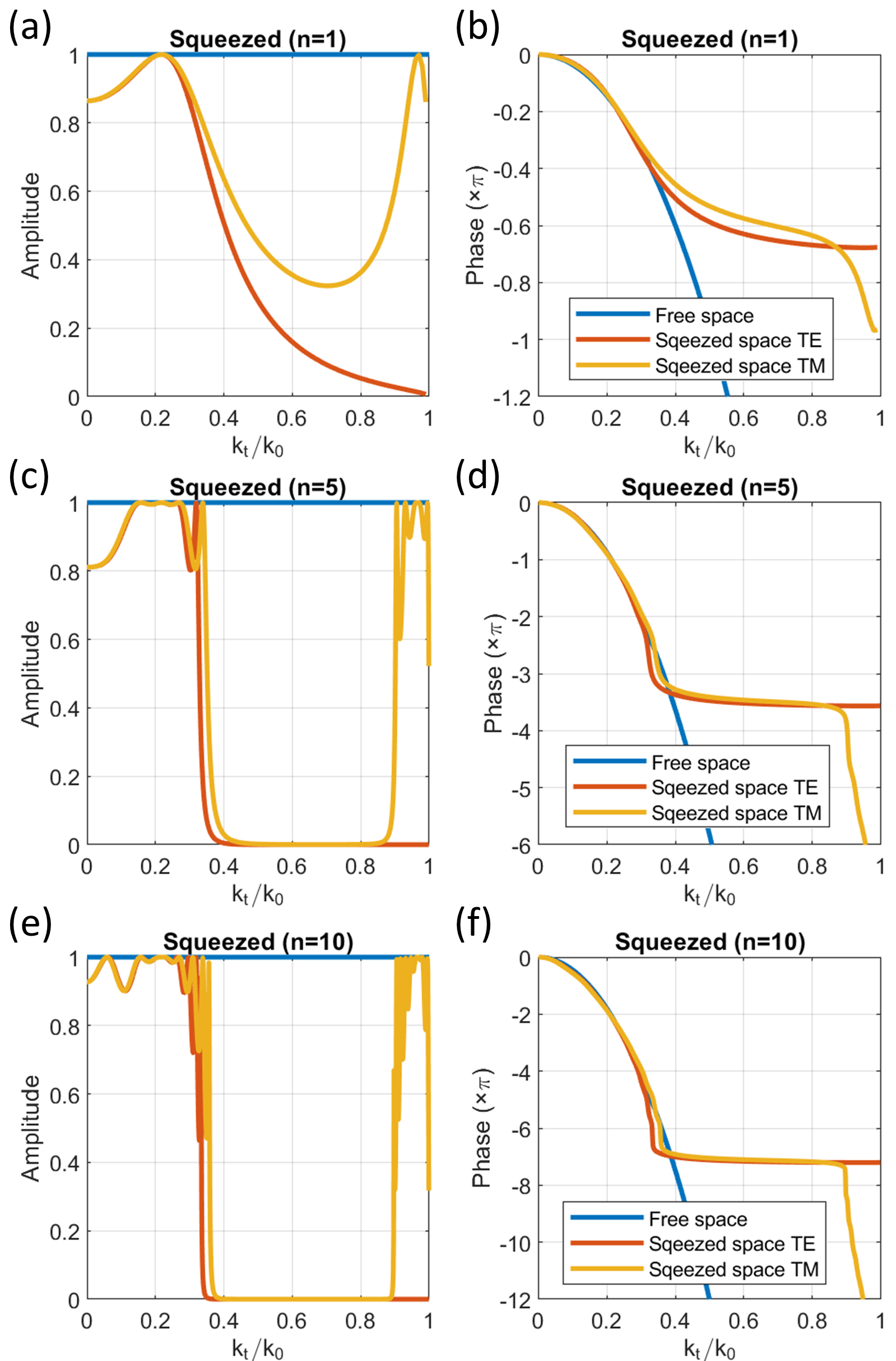}
	\caption{\label{fig4}Transmission amplitude (left column) and phase (right column) of the proposed $n$-layer structure acting as a nonlocal metasurface, for both TE and TM polarizations, compared with the transmission response of the replaced free-space volume of length: $3.6\lambda_0$ (first row), $21.8\lambda_0$ (second row) and $44.9\lambda_0$ (third row). The transmission coefficient is plotted as a function of transverse wavenumber $k_t$, at a fixed near-resonance frequency $\omega_0 = 1.02 \omega_r$. All transmission phases are adjusted to start from 0 at $k_t=0$ since a global phase difference with respect to free space is irrelevant. }
\end{figure}

Since the transmission phase is quadratic with $k_t$ at frequencies both slightly above and below resonance (where the phase is linear with respect to frequency before saturating), we operate slightly off-resonance, setting the operating frequency to $\omega_0=1.02\omega_r$, so that we can utilize a wider range of $k_t$ over which the phase is quadratic, consistent with our discussion around Eq. (\ref{eq:trade-off}) (this can also be understood by tracking the phase along the horizontal dashed lines in Fig. \ref{fig2}). The downside of this choice is that the transmission amplitude is no longer unity for $k_t=0$, as seen in Fig. \ref{fig2}; however, if we are not too far away from resonance, the amplitude remains high, especially if multiple layers are employed.
%which is a conservative choice, because the constant term in Eq.(\ref{eq:phase metasurface}) is around $0.2\pi$, far from the limit $0.5\pi$. However, this choice maintains a low transmission phase error and high transmission amplitude.
This is further demonstrated in Fig. \ref{fig4}, which shows the transmission amplitude and phase of the $n$-layer nonlocal structure as a function of transverse momentum $k_t$, at the operating frequency $\omega_0$. 
The analytical results are calculated using the standard transfer-matrix method. We observe that the $k_t$-dependent transmission phase of these nonlocal structures matches that of free space, with high accuracy, over a moderately wide angular range $-0.33<k_t/k_0<0.33$. This means that, with for example $n=10$ layers, corresponding to a thickness $T\approx 9\lambda_0$, we can replace a length of free space $L\approx 45\lambda_0$ over a reasonably large angular range. These numbers correspond to approximately 50\% of the limit in Eq. (\ref{eq:trade-off-n}), indicating good utilization of the available phase range. Furthermore, within this operating angular range, the transmission amplitude remains large, with a minimum of 0.8 for the single-resonator metasurface in Fig. \ref{fig4}(a) (this is much higher than the transmission amplitude of the nonlocal metamaterial in Ref. \cite{reshef2020towards}). We also note that the transmission response does not change much with polarization within the operating range of $k_t$. This is because (i) the transmission coefficients for TE and TM polarizations converge at normal incidence, and (ii) for moderately small incidence angles, the TE and TM transmission peaks are almost identical since the Fabry-Pérot-like resonances only depend on the phase delay inside the resonators, which is polarization independent, and on the reflection coefficients of the dielectric plates, which in our case remain approximately real for moderately small $k_t$. 

Most importantly, the total transmission phase range provided by the $n$-layer nonlocal structure keeps increasing linearly with $n$ (see the different phase ranges in Fig. \ref{fig4}(b,d,f)), with no observable increase of relative phase error and with clear improvements in terms of transmission amplitude. Since these multilayered nonlocal structures are able to mimic the response of free space for an arbitrary range of transmission phases, they can be used to replace free-space volumes of arbitrary length, as further demonstrated in the context of focusing and imaging in the next section. 

We also note that, because the replaced length of free space increases linearly with $n$, but the thickness of the $n$-layer nonlocal structure does not, due to the extra quarter-wavelength space between resonators, the squeezing ratio $R$ is not constant. For our one-layer nonlocal metasurface, the squeezing ratio is $R=5.6$, while it slightly decreases to $R=5.15$ when $n$ is large (see also supporting information). This compression ratio and angular range are better than the ones reported in Ref. \cite{reshef2020towards} ($R=4.9$ and $k_{t,\max}/k_0=0.27$), even without optimization, and with a much simpler, regular, and scalable design with higher transmission amplitude. In addition, if we target a much smaller angular range, for example $k_{t,\max}/k_0=0.01$, using a 40-layer structure, as in one of the designs in Ref. \cite{guo2020squeeze}, the total length of replaced free space $L/\lambda_0$ allowed by the trade-off in Eq. (\ref{eq:trade-off-n}) would be more than fifty times larger than what was obtained in Ref. \cite{guo2020squeeze}, due to the fact that these earlier designs only utilize a few percents of the available phase range around a transmission resonance (compared to 50\% for our structures). %\blue{(I think the advantage of our work is the unlimited length of replaced free space and large range of $k_t$, with simple, flexible and scalable structure. So only our device can do truly solid-state focusing and imaging. And we make full use of the phase range. The problem in \cite{guo2020squeeze} is mainly that the range of $k_t$ is too limited and the use of phase range is very inefficient. The problem of \cite{reshef2020towards} is lack of physical insight and not flexible and difficulty of scaling. So the total length of replaced free space is limited by the bound.)}
We stress again that our results are obtained with only dielectric materials and no optimization, which suggests that there is very large room for improvement in terms of compression ratio (for a fixed $L$ and $k_{t,\max}$) if the resonators are miniaturized using optimization and inverse design, and perhaps by employing high-index dielectrics combined with plasmonic materials.

\section{Nonlocal Flat-Optics for Focusing, Imaging, and \\Magnification}

\begin{figure}[htb]
	\includegraphics[ width=0.55\linewidth, keepaspectratio]{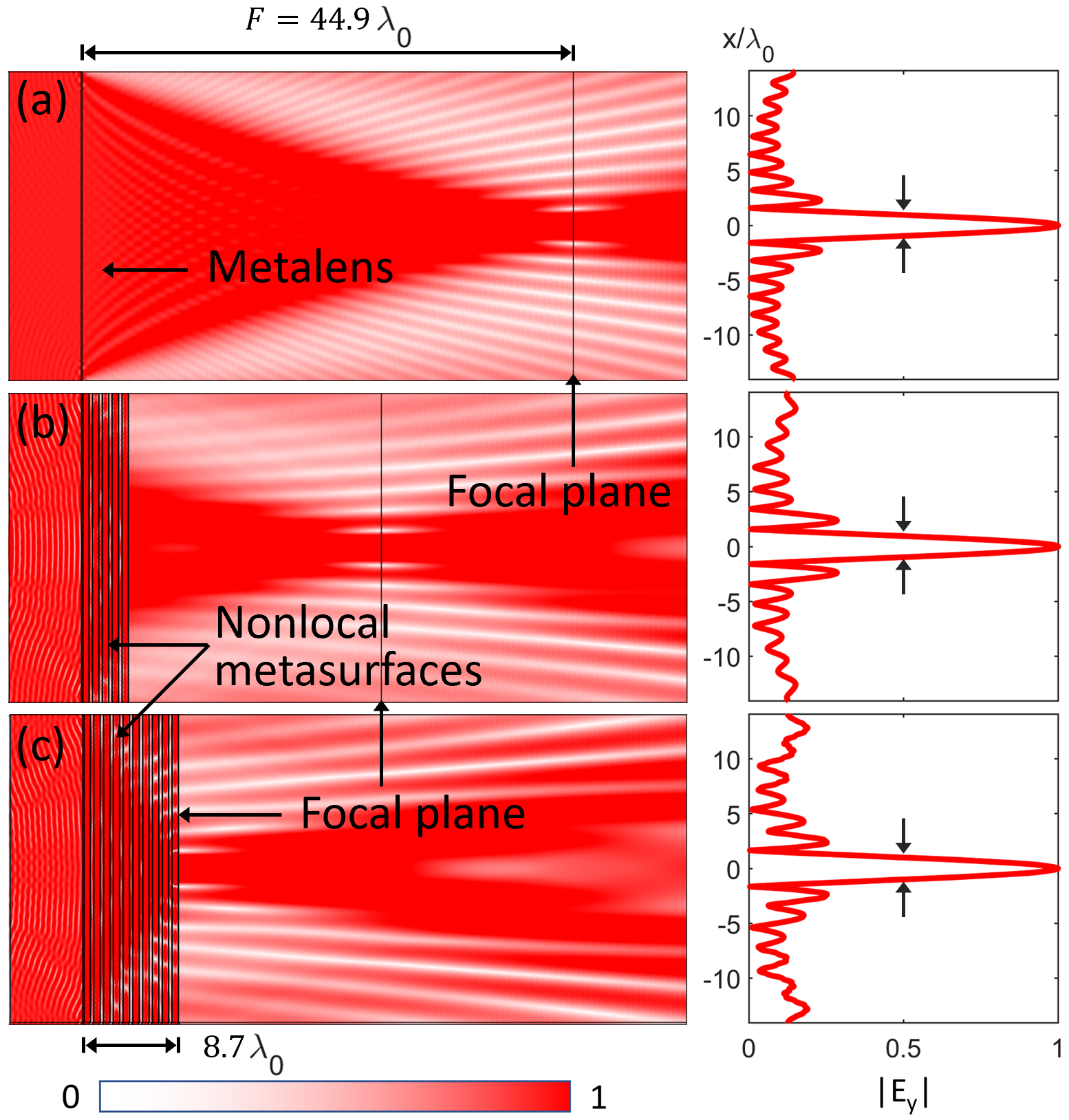}
	\caption{\label{fig5} Normalized distribution of the electric field amplitude for the focusing of a TE-polarized plane wave by (a) a local idealized metalens, and (b,c) the same metalens followed by nonlocal metasurfaces with 5 layers (panel b) and 10 layers (panel c). The dielectric nonlocal structure moves the focal plane closer and closer to the metalens, with minimal distortions and without changing the focal length of the metalens (which is a property of the local metalens itself). The insets on the right show the field amplitude distribution on the focal plane. In panel (c), all space between the lens and the focal plane has been replaced and compressed by the nonlocal structure, realizing a compact solid-state flat-optics focusing system.}
\end{figure}

In this section, we theoretically demonstrate, for the first time, the potential of combining local metalenses and nonlocal space-squeezing metasurfaces for focusing and imaging. First, we apply these ideas to the problem of focusing a TE-polarized plane wave within a much more compact volume. The results are shown in Fig. \ref{fig5} (obtained via full-wave numerical simulations \cite{COMSOL}). Because local metalenses have been studied extensively in the literature and the design of a specific metalens is not the topic of this paper, we use an idealized metasurface that implements the local transmission-phase response in Eq. (\ref{eq:1}) (an idealized planar thin slab with a suitable distribution of constitutive parameters); however, this may be replaced by any of the realistic metalenses demonstrated in the recent literature. We have also verified that, if the local metalens is not ideally impedance-matched to the surrounding medium, the resulting reflections and interactions between metalens and nonlocal metasurface reduce the overall transmission efficiency but, even in the presence of moderately high reflections, the functionality of the system is not affected too negatively (see supplementary information). This is mainly because the designed nonlocal structure exhibits very low reflection within its operating bandwidth and angular range (see, e.g., Fig. \ref{fig4}(e)), therefore minimizing any unwanted interaction with a mismatched metalens. 

As shown in Fig. \ref{fig5}(a), when the space beyond the metalens is occupied by free space, the transmitted plane wave is focused on a focal plane at a distance equal to the focal length $F$ of the local metalens, as for any conventional thin lens. On the focal plane, the field profile largely follows a sinc function, as expected for this two-dimensional focusing setup, with a nearly diffraction-limited focal spot of width $1.67\lambda_0$, for a numerical aperture $NA=0.3$. Then, we insert a 5-layer and a 10-layer structure acting as nonlocal space-compression metasurfaces as in Fig. \ref{fig1}, between the lens and the focal plane. The results shown in Fig. \ref{fig5}(b,c) clearly demonstrate that the entire field distribution behind the lens is shifted towards the lens by a distance $L-T$, and with low distortions, leading to a drastic reduction of the focal plane distance from the metalens. Particularly striking is the case with the 10-layer nonlocal device, in which light is focused on the back face of the stack of local and nonlocal structures. This means that all free space between the lens and the focal plane has been replaced and compressed, realizing a compact, planar, solid-state, focusing device. Furthermore, comparing the field distributions at the focal plane for these three cases in Fig. \ref{fig5}, we note that the width of the main lobe and the position of the first zeros are nearly identical, suggesting good focusing performance with small monochromatic aberrations. This is obtained thanks to the good angular transmission response of the nonlocal structure (see Fig. \ref{fig4}), which closely mimics the response of a free-space volume both in amplitude and phase. %Although some distortions of the side lobes are observed, because of phase errors and amplitude variations, the amplitude of side lobes are confined within similar levels. 

\begin{figure}[htb]
	\includegraphics[ width=0.7\linewidth, keepaspectratio]{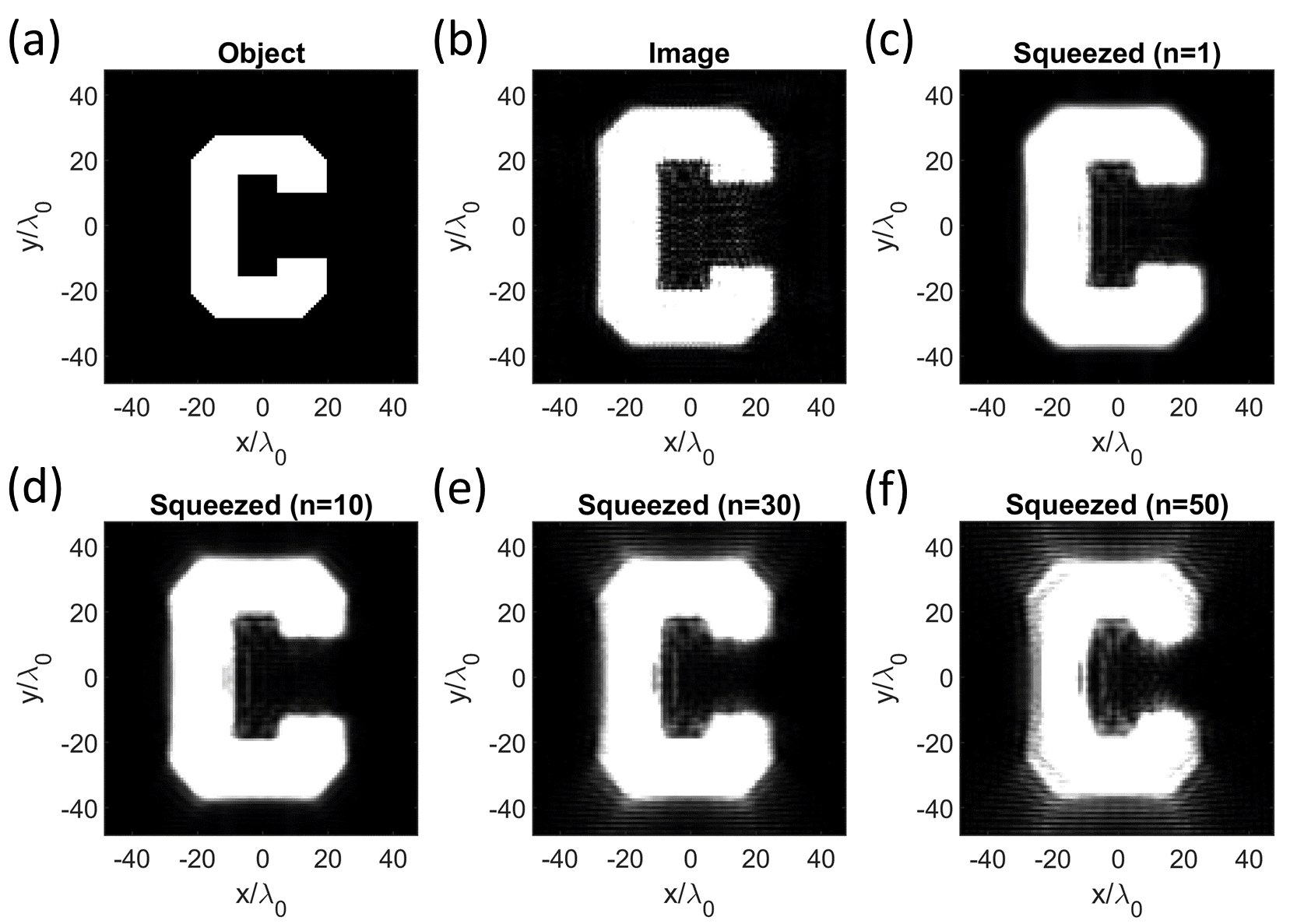}
	\caption{\label{fig6} Imaging and magnification with dielectric nonlocal metasurfaces inserted between a metalens and the observation plane (as illustrated in Fig. \ref{fig1} (a)). (a) Object to be imaged: a uniformly illuminated aperture with the shape of the letter ``C'' from Cornell University's logo. (b) Magnified image intensity, at the observation plane, formed by a metalens without nonlocal metasurfaces. (c)-(f) Magnified images formed by the same metalens followed by multilayered structures acting as nonlocal space-squeezing metasurfaces, as in Fig. \ref{fig5}, for different numbers of layers $n$. The same magnification is obtained in all panels since the focal length of the metalens remains the same, whereas the image forms at closer distances, thereby overcoming the trade-off between propagation length and magnification, as discussed in the text. The distance between lens and observation plane in panels (b) to (f) is, respectively, $230.6\lambda_0$, $226.6\lambda_0$, $193.4\lambda_0$, $119\lambda_0$ and $44.6\lambda_0$. The focal length of the metalens is 100$\lambda_0$. The considered lateral size of the object, flat-optics device, and observation plane is $96\lambda_0\times 96\lambda_0$. All of the transverse planes are discretized into $120\times 120$ pixels, hence the size of each pixel is $0.8\lambda_0\times 0.8\lambda_0$. In panel (f), all space between the lens and the observation plane has been replaced by the designed dielectric nonlocal device. %The distance between the lens and the observation plane is the length of replace free space by 50-layer metasurfaces, so in (f), all free space between the lens and the observation plane has been replaced metasurfaces (true solid-state optics).
	}
\end{figure}

Most importantly, we stress that, while focusing is obtained at a much shorter distance, as seen in Fig. \ref{fig5}(b,c), the focal length of the lens is the same (since the nonlocal structure is transversely homogeneous it cannot change the focal length of the metalens). Thus, the nonlocal device fundamentally decouples the actual distance at which focusing is achieved from the focal length of the lens, and hence its magnifying power. Indeed, this hybrid system combining local and nonlocal metasurfaces may realize magnification of an object without the need for a large propagation length. Fig. \ref{fig6} shows an example demonstrating this exciting opportunity for imaging and magnification. The object to be imaged is an illuminated aperture with the shape of the letter ``C'' in Cornell University's logo. We calculated the resulting image, produced by a composite flat-optics device as in Fig. \ref{fig5}(b,c), using the well-established scalar diffraction theory \cite{goodman2008introduction}. We assume the electric field is uniform in the $y$ direction on the aperture in Fig. \ref{fig6}(a) and zero outside (note that this assumption is not physical as it violates the continuity of tangential electric field, but the results under this assumption are in good agreement with experimental measurements for sufficiently large objects \cite{goodman2008introduction}; for separation of TE and TM polarizations, we refer to Ref. \cite{PhysRevLett.121.173004} (Supplemental material)). The imaging system is set up to obtain a magnified image, with a magnification factor of 1.3 (but larger values can be easily obtained by playing with the object distance and the focal length of the metalens). The image produced by the metalens alone, without nonlocal metasurfaces, is shown in Fig. \ref{fig6}(b) at a distance ($230.6 \lambda_0$) dictated by the focal length of the lens and the desired magnification factor. Then, as in Fig. \ref{fig5}, we insert the designed nonlocal structure, with increasing number of layers, right after the metalens. The resulting images, shown in Fig. \ref{fig6}(c-f), form on planes that are closer and closer to the metalens. When the inserted number of layers reaches 50, all free space is replaced by the much thinner nonlocal device, and the image forms on the back face of the structure. The distance between lens and observation plane has been decreased from $230.6 \lambda_0$ to $44.6 \lambda_0$; however, since the focal length of the metalens is unaltered, the same magnification is obtained as in the original case. 

In the results reported in Fig. \ref{fig6}, we observe larger distortions when $n$ is large because, although the relative phase error (with respect to the maximum transmission phase for a certain $n$) remains approximately constant, the absolute error increases with $n$. Then, when the absolute error is comparable with an odd-multiple of $\pi$, maximum distortion occur on the observation plane. However, the image is still clearly visible and, thanks to the high transmission amplitude of the proposed nonlocal structure, the image intensity remains high even for a cascade of 50 layers. We also note that the functionalities discussed in this section, in particular the fully solid-state focusing and imaging systems in Figs. \ref{fig5}(c) and \ref{fig6}(f), would not be realizable with earlier nonlocal designs as they either exhibit an angular range that is too narrow \cite{guo2020squeeze} or they are not scalable to many layers without further optimization, or drastic design changes, to obtain a high transmission amplitude \cite{reshef2020towards}. Despite its simplicity, our proposed design is the first to enable a much wider operating angular range and to offer the possibility to replace, in a scalable manner, an arbitrarily long free-space volume with high transmission amplitude. Thanks to these advances, we have theoretically demonstrated, for the first time, the potential of suitably combining local and nonlocal flat optics for focusing and imaging beyond some of the trade-offs of conventional optics.

\section{Conclusion}
In summary, we have proposed a solution to realize fully solid-state ultra-thin optical systems in which all conventional lenses are replaced by flat local metalenses and all free-space volumes between lenses and detectors are replaced by the proposed dielectric structures acting as nonlocal metasurfaces. From an analysis of the response of generic single-resonance-based nonlocal metasurfaces, we have derived a quantitative trade-off between the total length of replaced free space and the operating range of transverse momentum (angular range). This trade-off explains some of the inherent limitations and constraints of previous works \cite{reshef2020towards,guo2020squeeze}. To relax this trade-off and realize nonlocal space-compression optics with performance more suitable for practical applications, we have designed and theoretically/computationally demonstrated a nonlocal flat-optics structure composed of $n$ resonant layers separated by quarter-wavelength spacers, inspired by coupled-resonator-based microwave band-pass filters. This design strategy guarantees high transmission amplitude and highly linear transmission phase vs. frequency, without compromising the operating range of transverse momentum, hence enabling the possibility of replacing free-space volumes of arbitrary length. With our proposed design, we have demonstrated, theoretically and computationally, that nonlocal flat optics may allow realizing planar ultra-thin structures for light focusing and imaging that overcome some of the limits and trade-offs of conventional imaging systems, with intriguing implications for photography, augmented/virtual reality, microscopy, et cetera. 

%\red{[We also note that if the frequency of operation is varied, the device can still work, but within a different angular range, not necessarily centered at 0, but around a different value as can be understood by drawing a straight line in the dispersion diagrams of Fig. 2. Multi-wavelegth devices could also be envisioned if more dispersion branches are brought closer along the frequency axis...]}

The proposed nonlocal metasurfaces are based on a simple stack of homogeneous dielectric slabs, in which the high-permittivity layers could be made of silicon, separated by layers made of a transparent material with lower permittivity (for example silica). However, different materials could be considered as long as a moderately high refractive-index contrast between layers is obtained (the contrast would affect the linewidth of the resonances, and hence the replaced length of free space). We stress that since the $Q$ factor of the resonant layers is modest ($Q = 15$ in the considered example), the operating bandwidth of the nonlocal device is not too narrow (fractional bandwidth of few percent) and the device is expected to be moderately robust to losses, fabrication tolerances, and resonator detuning. These properties make our design particularly suitable for fabrication and integration between a metalens and an imaging sensor. Indeed, the fabrication of multi-layer dielectric optical coatings is a mature technology \cite{macleod2017thin} that can be leveraged for the future realization of these devices. We also would like to reiterate that the proposed structures have been obtained without any optimization, and hence we are confident there is still very large room for improvement, especially in terms of miniaturization of the coupled resonators composing the nonlocal device. 

Besides optical applications, our proposed nonlocal metasurfaces may also be used at RF and microwave frequencies, e.g., for antenna miniaturization, and an experimental proof-of-concept in this area is currently under way. More broadly, we strongly believe that the vast opportunities offered by combining local and nonlocal metasurfaces may have far-reaching applications in different areas of wave physics and engineering.

\section*{Funding}

The authors acknowledge support from the Air Force Office of Scientific Research with Grant No. FA9550-19-1-0043 through Dr. Arje Nachman, and the National Science Foundation (NSF) with Grant No. 1741694.

\section*{Supporting Information}
The Supporting Information is available free of charge via the internet at http://pubs.acs.org. Approximate analytical models for nonlocal metasurfaces; Approximate expressions for the dispersion relation and the length of replaced free space; Impact of impedance mismatch and reflections. (PDF)

% Bibliography
\bibliography{acs-achemso}

\end{document}